\documentclass[pra,twocolumn,showpacs,preprintnumbers,amsmath,amssymb,superscriptaddress]{revtex4-1}

\usepackage{graphicx}
\usepackage{dcolumn}
\usepackage{bm}
\usepackage{comment}
\usepackage{bbold}

\begin{document}

\title{Tachyon physics with trapped ions}
\date{\today}

\author{Tony E. Lee}
\affiliation{ITAMP, Harvard-Smithsonian Center for Astrophysics, Cambridge, Massachusetts 02138, USA}
\affiliation{Department of Physics, Harvard University, Cambridge, Massachusetts 02138, USA}
\author{Unai Alvarez-Rodriguez}
\affiliation{Department of Physical Chemistry, University of the Basque Country UPV/EHU, Apartado 644, 48080 Bilbao, Spain} 
\author{Xiao-Hang Cheng}
\affiliation{Department of Physical Chemistry, University of the Basque Country UPV/EHU, Apartado 644, 48080 Bilbao, Spain}
\affiliation{Department of Physics, Shanghai University, 200444 Shanghai, People's Republic of China}
\author{Lucas Lamata}
\affiliation{Department of Physical Chemistry, University of the Basque Country UPV/EHU, Apartado 644, 48080 Bilbao, Spain}
\author{Enrique Solano}
\affiliation{Department of Physical Chemistry, University of the Basque Country UPV/EHU, Apartado 644, 48080 Bilbao, Spain}
\affiliation{IKERBASQUE, Basque Foundation for Science, Maria Diaz de Haro 3, 48013 Bilbao, Spain}

\begin{abstract}
It has been predicted that particles with imaginary mass, called tachyons, would be able to travel faster than the speed of light. There has not been any experimental evidence for tachyons occurring naturally. Here, we propose how to experimentally simulate Dirac tachyons with trapped ions. Quantum measurement on a Dirac particle simulated by a trapped ion causes it to have an imaginary mass so that it may travel faster than the effective speed of light. We show that a Dirac tachyon must have spinor-motion correlation in order to be superluminal. We also show that it exhibits significantly more Klein tunneling than a normal Dirac particle. We provide numerical simulations of realistic ion systems and show that our scheme is feasible with current technology. 
\end{abstract}

\maketitle

\section{Introduction}

In the 1960s, it was theorized that there could be particles with imaginary mass, called tachyons, which travel faster than light \cite{bilaniuk62,feinberg67,bilaniuk69,aharonov69,chodos85}. This is because in order for relativistic energy $E=mc^2/\sqrt{1-(v/c)^2}$ to be real when $m$ is imaginary, $v$ must be larger than $c$. Note that although a tachyon is considered to be superluminal, causality may still be preserved \cite{aharonov69,chiao96}. There has been no experimental evidence of tachyons occurring naturally \cite{alvager68,baltay70}. There have been a few proposals for engineering tachyon-like excitations that travel faster than the effective speed of light in the system. These include inverted optical media \cite{chiao96} and waveguides \cite{szameit11,longhi11,marini14}. The motivation for an experimental realization is that it would broaden the range of accessible phenomena, and allow the further study of physics that would otherwise be unphysical.


Ion traps have proven to be an ideal setting for experimentally realizing quantum relativistic effects \cite{lamata07,casanova10,casanova11,loh13,gerritsma10,gerritsma11,alvarez13,zhang15,lamata11}. Other platforms have been considered as well \cite{otterbach09,salger11,keil14}. Recent trapped-ion experiments have simulated Dirac particles and observed \emph{Zitterbewegung} and Klein tunneling \cite{gerritsma10,gerritsma11}. In these experiments, the excellent control and read-out capabilities allow one to prepare and monitor the wavepacket dynamics.

\begin{figure}[b]
\centering
\includegraphics[width=3.4 in,trim=1.2in 4.2in 1in 4.1in,clip]{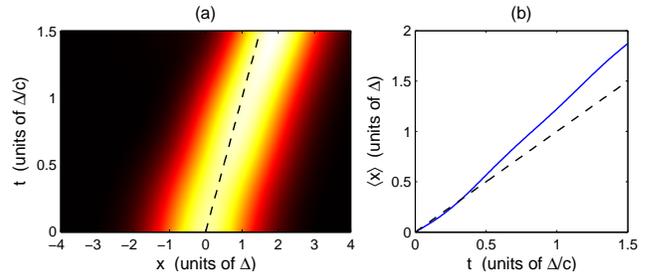}
\caption{\label{fig:free} (color online) Propagation of a tachyon wave packet. (a) Probability density $|\psi(x,t)|^2$ is plotted on color scale. Dashed line denotes light cone. (b) Solid line is average position $\langle x\rangle$ as a function of $t$. Dashed line denotes speed of light. Initial wavefunction is $\psi(x)=e^{ip_ox}e^{-x^2/4\Delta^2}u_+(p_o)$ with $p_o=3.5/\Delta$ and $m=2/(c\Delta)$. }
\end{figure}

In this paper, we propose a scheme to simulate Dirac tachyons \cite{lemke76} with trapped ions and show that their quantum nature distinguishes them from classical tachyons. In our scheme, continuous measurement on a Dirac particle simulated by a trapped ion causes it to have an imaginary mass. The Dirac tachyon can then move faster than the effective speed of light in the system (see Fig.~\ref{fig:free}). We perform realistic numerical simulations with example experimental numbers and show that this scheme is feasible with current technology. We also describe how to measure the relevant observables.

We also obtain new results regarding the properties of Dirac tachyons. We show that spinor-motion correlation plays a crucial role in the propagation, i.e., there must be spinor-motion correlation in order for a tachyon to be superluminal. We also consider the interaction with an external potential and find that a tachyon exhibits significantly more Klein tunneling than a normal particle. Then, we use a spacetime duality to show that a normal particle scattering off a spatially-varying potential is dual to a tachyon scattering off a time-varying potential.  All these features can be simulated with trapped ions following our proposal.

Our paper is structured as follows. In Sec.~\ref{sec:model}, we introduce the Dirac equation with imaginary mass and describe our proposed experimental scheme. In Sec.~\ref{sec:velocity}, we discuss the velocity of the wavepacket and its relationship to entanglement. In Sec.~\ref{sec:scattering}, we calculate Klein tunneling for a Dirac tachyon. In Sec.~\ref{sec:experiment}, we provide experimental numbers and comment on the scalability of our scheme.

\section{Model} \label{sec:model}

A quantum relativistic particle is described by the Dirac equation \cite{thaller92}. In one spatial dimension, it takes the form $i\partial_t\psi=H\psi$, where \cite{thaller04}
\begin{eqnarray}
H=cp\sigma_x+mc^2\sigma_z \label{eq:dirac_normal}
\end{eqnarray}
with speed of light $c$, mass $m$, momentum operator $p$, and Pauli matrices $\sigma_k$. We assume that $\hbar=1$. The wavefunction $\psi$ has both motional and spinor degrees of freedom. For the two-component spinor, we work in the $\sigma_z$ basis: $\left|\uparrow\right\rangle$ and $\left|\downarrow\right\rangle$.

This equation was experimentally implemented with a trapped ion by letting the ion's electronic and vibrational degrees of freedom correspond to a Dirac particle's spinor and motional degrees of freedom \cite{lamata07,gerritsma10}. The term $p\sigma_x$ can be written in terms of $a\sigma_\pm$ and $a^\dagger\sigma_\pm$, so by exciting both blue and red motional sidebands, one obtains Eq.~\eqref{eq:dirac_normal}. The effective speed of light and mass are $c=2\eta\Delta\tilde{\Omega}$ and $mc^2=\Omega$, where $\eta$ is the Lamb-Dicke parameter, $\Delta$ is the size of the ground-state wavefunction, $\tilde{\Omega}$ is the strength of the carrier transition, and $\Omega$ is the detuning.

Now suppose $\left|\uparrow\right\rangle$ has a finite lifetime given by a linewidth $\gamma$. Then, conditioned on the absence of a decay event, the system evolves with a non-Hermitian Hamiltonian \cite{dalibard92,molmer93,dum92,daley14},
\begin{eqnarray}
H&=& H_o - \frac{i\gamma}{4}\sigma_z. \label{eq:Heff}
\end{eqnarray}
This is because the null measurement of a decay event has back-action on the wavefunction, which is accounted for by the non-Hermitian term. (See footnote \footnote{The non-Hermitian term is usually written as a projector $\sigma^+\sigma^-=\sigma_z/2 + 1/2$. The additional constant $1/2$ only affects normalization: since the wavefunction is normalized when calculating expectation values, this constant does not affect the physics. So we have omitted the constant in Eq.~\eqref{eq:Heff}}).  Letting $H_o=cp\sigma_x$, we then have the Dirac equation with imaginary mass \cite{lemke76},
\begin{eqnarray}
H&=&cp\sigma_x-imc^2\sigma_z, \label{eq:dirac_tachyon}
\end{eqnarray}
where $mc^2=\gamma/4$. [Note that the sign of the mass term in Eq.~\eqref{eq:dirac_tachyon} is unimportant since it can be flipped via a unitary transformation.] It is important to condition on the absence of decay events; if decay events were included, the dynamics would be described by a master equation instead of a non-Hermitian Hamiltonian \cite{lee13b,zou14,hush14}. Thus, quantum measurement allows one to simulate a Dirac tachyon with a trapped ion.

The experimental protocol is as follows. Starting from an initial wavepacket, one evolves the ion with $H_o=cp\sigma_x$, by resonantly exciting blue and red sidebands, while optically pumping $\left|\uparrow\right\rangle$ to an auxiliary state. After a given amount of time, one turns off $H_o$ and the optical pumping, and then measures the population in the auxiliary state using the usual fluorescence method. If one measures no population in the auxiliary state, then there was no decay event and therefore the ion was evolving solely according to Eq.~\eqref{eq:dirac_tachyon}. This is a probabilistic but heralded method \cite{sherman13,lee14b,lee14d}: one repeats the protocol many times and post-selects on those runs without decay events, since those are the ones that simulate Eq.~\eqref{eq:dirac_tachyon}. In Sec.~\ref{sec:experiment}, we provide experimental numbers and numerical simulations with realistic decoherence and show that the probability of a successful run is reasonably high. Also, it is not necessary to efficiently detect a single photon; instead, one only needs to efficiently measure the population in the auxiliary state.

\begin{figure}[t]
\centering
\includegraphics[width=3.5 in,trim=1.in 3.9in 1in 3.9in,clip]{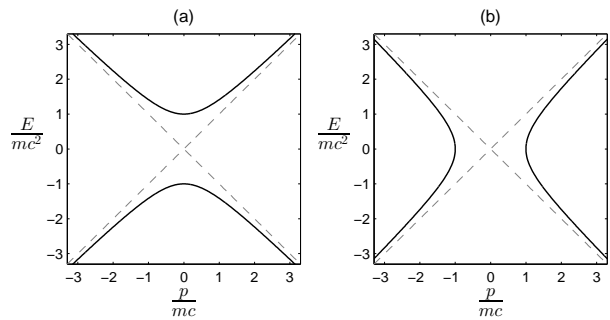}
\caption{\label{fig:dispersion}Dispersion relations (energy vs. momentum) for (a) normal massive particle and (b) tachyon. Dashed lines denote speed of light.}
\end{figure}

We distinguish our scheme from related work. If the second term of Eq.~\eqref{eq:dirac_tachyon} included a complex-conjugation operator, as in the Majorana equation \cite{casanova11,zhang15,keil14}, one could simulate it by embedding the spinor in a larger Hilbert space; however, this does not work for the tachyon quantum simulation.
Also, a Dirac tachyon is different from a Klein-Gordon tachyon \cite{aharonov69,chiao96} because of the spinor degree of freedom, which is important to the dynamics. In addition, our scheme is different from superluminality in an absorptive medium with anomalous dispersion \cite{garret70,chu82,steinberg93}, because the latter is due to pulse reshaping instead of imaginary mass and does not have a tachyonic dispersion [Fig.~\ref{fig:dispersion}(b)].

We now compare the behavior of normal massive particles [described by Eq.~\eqref{eq:dirac_normal}] with tachyons [described by Eq.~\eqref{eq:dirac_tachyon}].

\section{Velocity} \label{sec:velocity}

A normal massive particle travels slower than $c$, because the group velocity $v_g=dE/dp$ is smaller in magnitude than $c$, where $E=\pm\sqrt{p^2c^2 + m^2c^4}$ is the dispersion of Eq.~\eqref{eq:dirac_normal}, plotted in Fig.~\ref{fig:dispersion}(a). Another way to see this is from the equation of motion for the position operator, $\partial_t\langle x\rangle = c\langle\sigma_x\rangle$, which is bounded above by $c$.

The special property of a tachyon is that its group velocity is larger than $c$. This is because the dispersion of Eq.~\eqref{eq:dirac_tachyon} is $E=\pm\sqrt{p^2c^2 - m^2c^4}$, as shown in Fig.~\ref{fig:dispersion}(b).  Figure \ref{fig:free} is a spacetime plot of a tachyon, showing that the wave packet is indeed superluminal.

Consider the equation of motion of $\langle x\rangle$ for a tachyon. For a non-Hermitian Hamiltonian like Eq.~\eqref{eq:dirac_tachyon}, the equation of motion for an operator includes extra terms \cite{graefe08b}, leading to
\begin{eqnarray}
\partial_t \langle x\rangle=c\langle\sigma_x\rangle - 2mc^2(\langle x\sigma_z\rangle - \langle x\rangle\langle\sigma_z\rangle). \label{eq:dxdt}
\end{eqnarray}
Since the first term is bounded by $c$, in order to have $|\partial_t \langle x\rangle| > c$, the quantity in parentheses (correlation of $x$ and $\sigma_z$) must be nonzero. In other words, in order for a Dirac tachyon to be superluminal, the spinor and motional degrees of freedom must be correlated. Without correlations, a Dirac tachyon is actually subluminal. (For the case of a pure state, superluminality requires spinor-motion entanglement, but in an experiment, the state will be mixed.)

\begin{figure}[t]
\centering
\begin{tabular}{cc}
\includegraphics[width=1.6 in,trim=2.7in 4.in 2.9in 4.1in,clip]{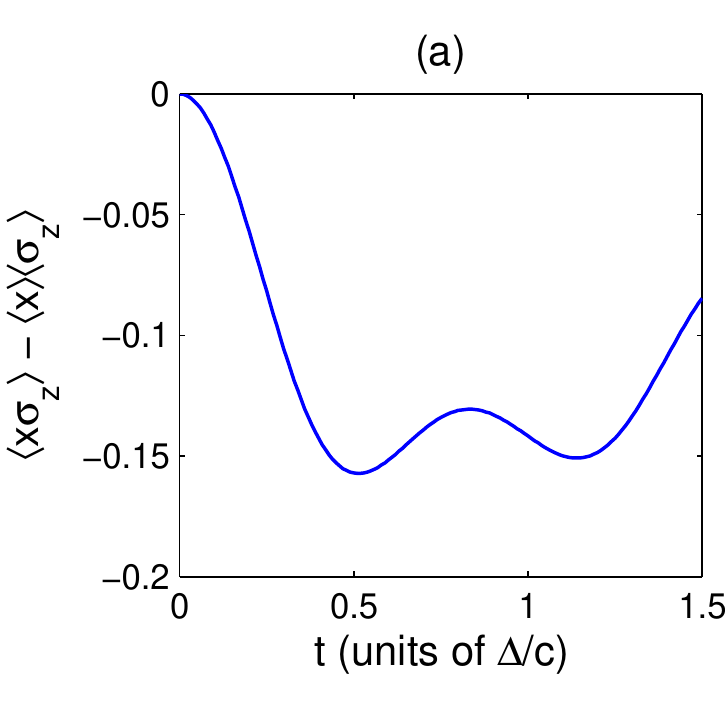}
\includegraphics[width=1.6 in,trim=2.7in 4.in 2.9in 4.1in,clip]{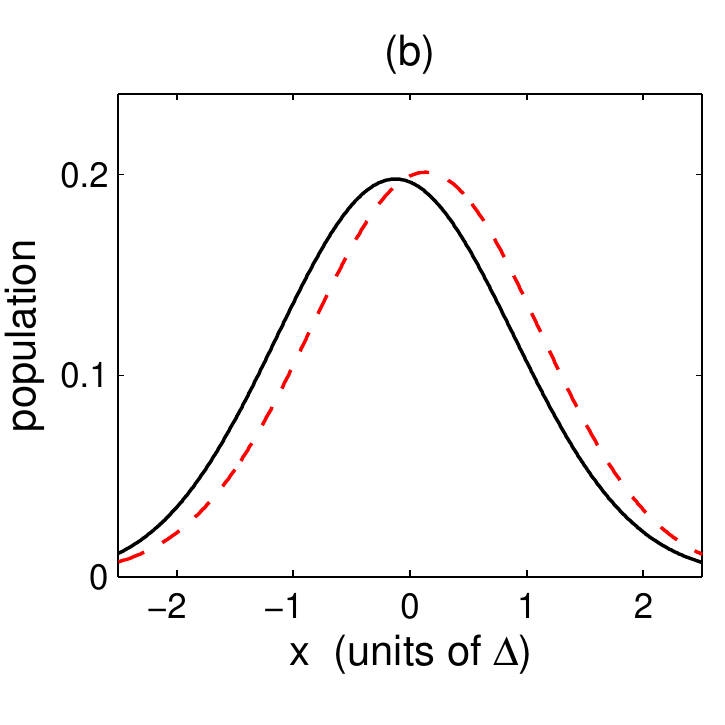}
\end{tabular}
\caption{\label{fig:entangled}(color online)  (a) Correlation of $x$ and $\sigma_z$ for Fig.~\ref{fig:free}(b). (b) Population in $\left|\uparrow\right\rangle$ (solid line) and $\left|\downarrow\right\rangle$ (dashed line) for the state in Eq.~\eqref{eq:entangled} with $p_o=3.5/\Delta$ and $m=2/(c\Delta)$.}
\end{figure}

This explains the behavior in Fig.~\ref{fig:free}(b). The initial wavefunction is $\psi(x)=e^{ip_ox}e^{-x^2/4\Delta^2}u_+(p_o)$, where the spinor $u_+(p_o)$ is the positive-energy eigenstate of Eq.~\eqref{eq:dirac_tachyon} for $p=p_o$. Since this is a separable state, the tachyon is initially subluminal. Over time, the wavefunction develops correlations [Fig.~\ref{fig:entangled}(a)] so that it eventually becomes superluminal. This oscillatory motion is an example of \emph{Zitterbewegung} \cite{thaller92,thaller04} and occurs because the initial wavefunction contains both positive and negative-energy components that interfere with each other. In momentum space, $\psi(p)$ is a Gaussian centered at $p_o$, but $u_+(p_o)$ is the positive eigenstate only for $p=p_o$.

The following entangled state,
\begin{eqnarray}
\psi(x)&=&\int dp\,e^{-\Delta^2(p-p_o)^2} e^{ipx} u_+(p), \label{eq:entangled}
\end{eqnarray}
contains only positive-energy components because the spinor $u_+(p)$ is the positive eigenstate for each $p$. Thus, this wavepacket travels superluminally with group velocity $v_g(p_o)$ and without \emph{Zitterbewegung}. For motion to the right, $x$ and $\sigma_z$ are negatively correlated, i.e., the peak of $|\psi_\downarrow(x)|^2$ is ahead of $|\psi_\uparrow(x)|^2$, as shown in Fig.~\ref{fig:entangled}(b). In the limit of $p_o\gg mc$, $\langle x\sigma_z\rangle - \langle x\rangle\langle\sigma_z\rangle=-mc/2p_o^2$. In contrast, for a normal particle in the state given by Eq.~\eqref{eq:entangled}, $\langle x\sigma_z\rangle - \langle x\rangle\langle\sigma_z\rangle=0$.

\section{Scattering off a potential} \label{sec:scattering}

\begin{figure}[b]
\centering
\includegraphics[width=3.5 in,trim=1.in 3.9in 1in 4.1in,clip]{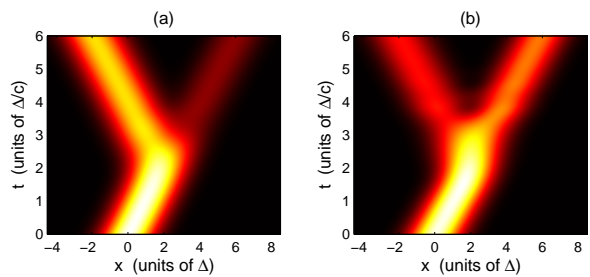}
\caption{\label{fig:phi} (color online)  Probability density $|\psi(x,t)|^2$ scattering off a linear electric potential for (a) normal massive particle and (b) tachyon. For both plots, $m=1/(c\Delta)$ and $g=2c/\Delta^2$.}
\end{figure}

Here, we consider the scattering off a repulsive electric potential that is linear in space, $e\phi(x)=gx$, where $e$ is the charge and $g>0$. It is well known that a normal Dirac particle can tunnel through such a barrier and propagate undamped due to the negative-energy branch of the dispersion [Fig.~\ref{fig:dispersion}(a)]. An incoming wavepacket splits into a reflected component and a tunneling component [Fig.~\ref{fig:phi}(a)], and the tunneling probability is $P_{tunnel}=\exp(-\pi m^2c^3/g)$ \cite{sauter31}. This is known as Klein tunneling \cite{klein29} and is surprising because a nonrelativistic Schr\"odinger particle would not tunnel without attenuation. In the trapped-ion implementation, a linear potential can be engineered by using a second ion, as demonstrated recently \cite{casanova10,gerritsma11}.

Now consider a tachyon scattering off a linear electric potential [Fig.~\ref{fig:phi}(b)],
\begin{eqnarray}
i\partial_t\psi&=&(cp\sigma_x-imc^2\sigma_z + gx)\psi. \label{eq:electric_potential}
\end{eqnarray}
Suppose an initial wavepacket moves to the right with $p=p_o>0$ and $E>0$. We work in momentum space and write $x=i\partial_p$ and $\psi(p,t)=\xi(p_o=p+gt,t)$ \cite{casanova10}:
\begin{eqnarray}
i\partial_t \xi=[c(p_o-gt)\sigma_x-imc^2\sigma_z]\xi. \label{eq:LZ}
\end{eqnarray}
Thus, the motion of the wavepacket is equivalent to a Landau-Zener process with a magnetic field $c(p_o-gt)$ that decreases linearly in time \cite{zener32}. The energy levels of this process are given by the dispersion in Fig.~\ref{fig:dispersion}(b). At $t=0$, the system starts at $p=p_o$ on the positive-energy branch. $p$ is then ramped to negative values, and the question is how the final population is distributed between the branches. The population in the positive-energy branch corresponds to the (left-moving) reflected component, while the population in the negative-energy branch corresponds to the (right-moving) tunneling component. [Note that the eigenvalues for $p\in(-mc,mc)$ are complex and not plotted in Fig.~\ref{fig:dispersion}(b)].

However, this is a non-Hermitian Landau-Zener process due to the imaginary field in Eq.~\eqref{eq:LZ}. By extending Zener's original solution \cite{zener32} to the non-Hermitian case, we find that the tunneling probability is
\begin{eqnarray}
P_{tunnel}&=&\frac{\exp\left(\frac{\pi m^2c^3}{g}\right)}{2\exp\left(\frac{\pi m^2c^3}{g}\right) - 1 }.\label{eq:prob_tunnel}
\end{eqnarray}
This result differs from that for a normal particle: the tunneling probability for a tachyon is \emph{larger} and is always at least 1/2. For example, Fig.~\ref{fig:phi} shows that a lot more population tunnels through for a tachyon than a normal particle. Physically, this is because the tachyonic dispersion does not have an energy gap, while the normal dispersion does (Fig.~\ref{fig:dispersion}).

There is a useful duality between normal Dirac particles and tachyons. Suppose $\psi(x,t)$ is a solution to the Dirac equation for a normal particle with electric potential $\phi(x)$ and vector potential $A(x)$ \cite{thaller92,casanova10}. One can show that $\psi(x',t')=U^{-1}\psi(x=t',t=x')$ is a solution to the Dirac equation for a tachyon with electric potential $\phi'(t')=-A(x=t')$ and vector potential $A'(t')=-\phi(x=t')$, where $U=(I+i\sigma_x)/\sqrt{2}$ is a unitary transformation. Thus, a normal particle scattering off a spatially-varying electric potential is dual to a tachyon scattering off a time-varying vector potential.

\section{Experimental considerations} \label{sec:experiment}

\subsection{Probabilities}

The tachyon protocol is probabilistic (i.e., conditioned on no decay event), so it is important to estimate the probability of a successful experimental run. To do this, we note that the average number of decay events during time $t$ is
\begin{eqnarray}
\mu&=&\frac{\gamma(\langle\sigma_z\rangle + 1)t}{2} \approx \frac{\gamma t}{2}, \label{eq:mu}
\end{eqnarray}
where we have used the fact that the eigenstates of Eq.~\eqref{eq:dirac_tachyon} satisfy $\langle\sigma_z\rangle=0$.  The number of decay events is a Poissonian random variable. Thus, the probability of a successful run (zero decay events during time $t$) is:
\begin{eqnarray}
P_\text{success}&=&e^{-\mu} \approx e^{-\frac{\gamma t}{2}}.
\end{eqnarray}
(To calculate $P_\text{success}$ exactly, one would use the instantaneous value of $\langle\sigma_z\rangle$, but we have found that setting $\langle\sigma_z\rangle=0$ in Eq.~\eqref{eq:mu} results in a very good estimate.)

We recall from Eq.~\eqref{eq:dirac_tachyon} that $\gamma = 4mc^2$. We also note that $m$ is in units of $1/(c\Delta)$, while $t$ is in units of $\Delta/c$ (since $\hbar=1$):
\begin{eqnarray}
m=m'\left(\frac{1}{c\Delta}\right), \quad t=t'\left(\frac{\Delta}{c}\right),
\end{eqnarray}
where $m',t'$ are dimensionless. Thus, 
\begin{eqnarray}
\mu\approx 2m't', \quad P_\text{success}\approx e^{-2m't'}.
\end{eqnarray}

As an example, consider Fig.~\ref{fig:free}, where $m'=2$. The particle is superluminal by $t'=0.5$, which corresponds to $P_\text{success}=e^{-2}\approx 0.14$. A time of $t'=1$ corresponds to $P_\text{success}=e^{-4}\approx 0.018$. These probabilities are reasonably high enough for the scheme to be feasible. $P_\text{success}$ can be easily increased by decreasing $m'$. Also, we note from Eq.~\eqref{eq:dxdt} that if the initial wavefunction is suitably entangled \cite{lo14}, the initial velocity can be larger than $c$, so that superluminality occurs earlier. For the case of an external potential, one can observe Klein tunneling with $P_\text{success}\approx 0.03$ by using a large $g$.

\subsection{Example numbers} \label{sec:example}

\begin{figure}[t]
\centering
\begin{tabular}{cc}
\includegraphics[width=1.7 in,trim=5in 1.9in 2.5in 2.3in,clip]{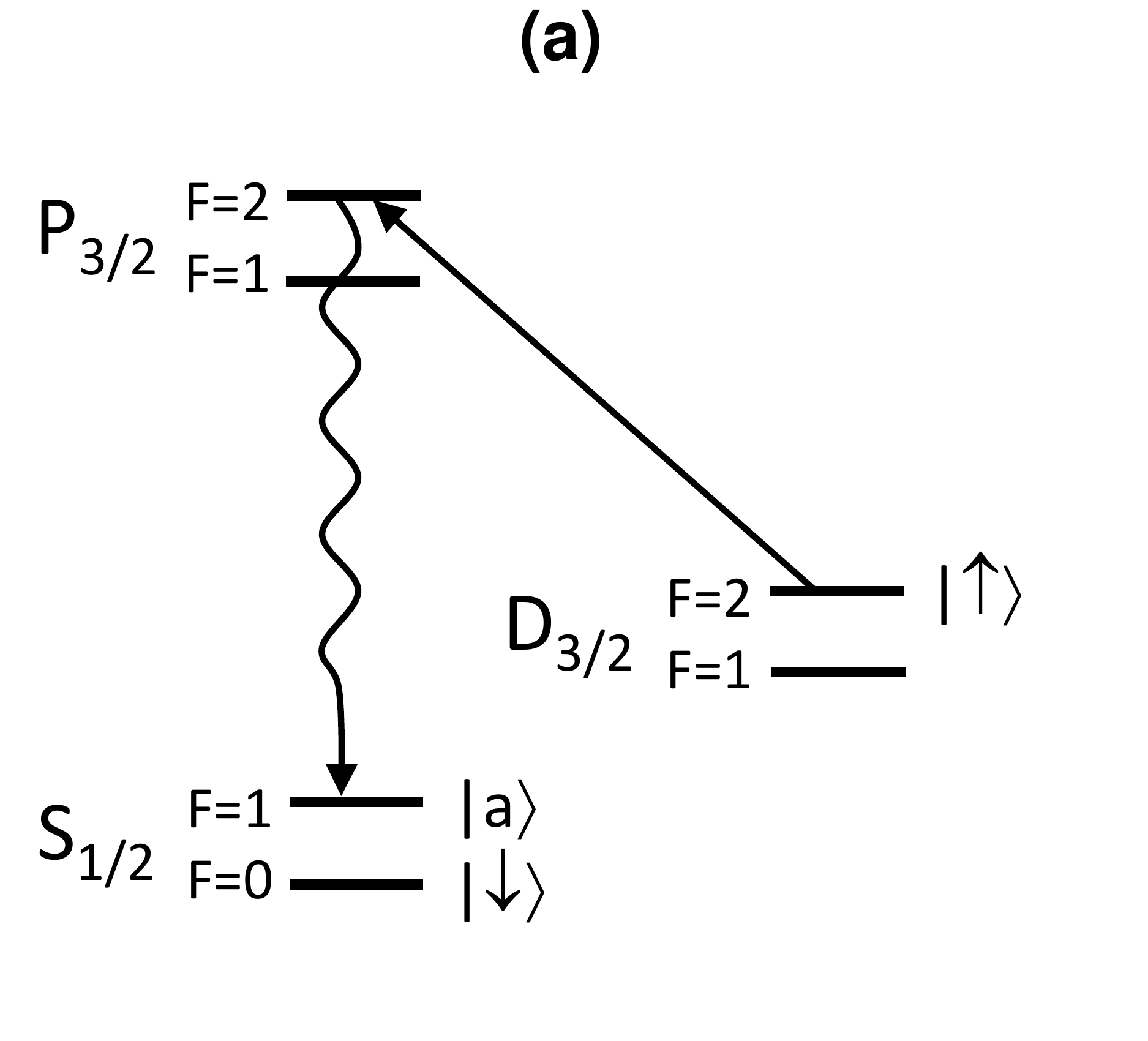}
\includegraphics[width=2.4 in,trim=2.5in 4in 0in 3.8in,clip]{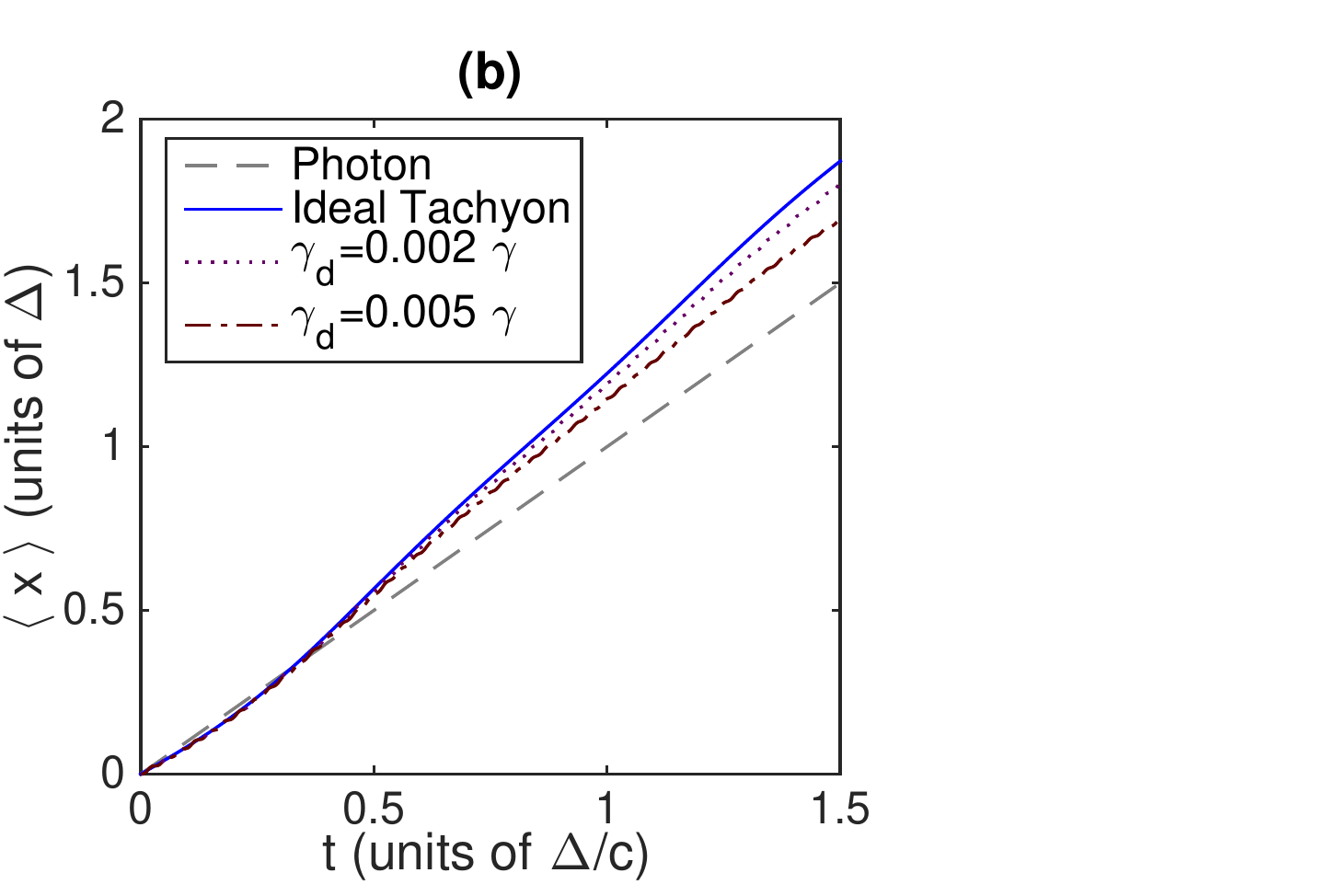}
\end{tabular}
\caption{\label{fig:experiment} (color online) (a) Level scheme for ${}^{171}\text{Yb}^+$. (b) Numerical simulation of trapped-ion implementation for different rates of optical-pumping errors.}
\end{figure}

We provide example numbers for ${}^{171}\text{Yb}^+$. See the level scheme in Fig.~\ref{fig:experiment}(a). To implement the first term of Eq.~\eqref{eq:dirac_tachyon}, one drives the red and blue sidebands of the $\left|\downarrow\right\rangle$--$\left|\uparrow\right\rangle$ transition (either directly or via a $P_{1/2}$ state). To implement the second term of Eq.~\eqref{eq:dirac_tachyon}, one optically pumps $\left|\uparrow\right\rangle$ to the auxiliary state $|a\rangle$ via the ${|P_{3/2},F=2\rangle}$ state. Due to dipole-selection rules, ${|P_{3/2},F=2\rangle}$ decays into $|a\rangle$ instead of $\left|\downarrow\right\rangle$. Also, the branching ratio of ${|P_{3/2},F=2\rangle}$ to $\left|\uparrow\right\rangle$ is less than $0.002$  \cite{biemont98,olmschenk09}, so these dephasing errors can be neglected [Fig.~\ref{fig:experiment}(b)]. To measure the population in $|a\rangle$, one excites $|a\rangle$ to $|P_{1/2},F=0\rangle$ and looks for fluorescence. To measure the ion position over time, one uses the procedure described in Refs.~\cite{gerritsma10,gerritsma11}.


Setting $\eta=0.05$, $\Delta=3.4~\text{nm}$, and $\tilde{\Omega}=2\pi\times 100~\text{kHz}$ means an effective speed of light $c=2\times 10^{-4}~\text{m/s}$ and time scale $\Delta/c=16~\mu\text{s}$. For optical pumping $\gamma=2\pi\times 80~\text{kHz}$, the mass is $m=2/(c\Delta)$. These numbers are readily accessible in current experiments.

\subsection{Numerical simulations} \label{sec:simulation}

We have performed numerical simulations of the tachyon protocol including realistic decoherence present in ion traps. The carrier term is eliminated by using two counterpropagating laser fields for red and blue sidebands, $H_o=i \tilde{\Omega}(\sigma_{-}e^{i \delta t} -\sigma_{+} e^{-i \delta t})[\sin \phi + \eta  \cos \phi (a e^{-i \nu t}+a^{\dag} e^{i \nu t}) ]$, where $\nu$ is the trap frequency and the laser phase $\phi$ is chosen such that $\sin \phi=0$. For the tachyon protocol, the intrinsic source of decoherence is optical pumping errors, i.e., when ${|P_{3/2},F=2\rangle}$ decays back to $\left|\uparrow\right\rangle$, which leads to dephasing with rate $\gamma_d$. Figure \ref{fig:experiment}(b) shows that the optical pumping errors are negligible for ${}^{171}\text{Yb}^+$, since $\gamma_d=0.002\gamma$. We have also found that since $\gamma_d$ is much larger than other sources of decoherence (motional heating, spontaneous emission, and laser dephasing) in current experiments, the latter have even less effect. Thus, the tachyonic behaviour is preserved with realistic decoherence.



\subsection{Dependence on detector efficiency} \label{sec:detector}

Since our scheme is conditioned on the absence of a decay event, one must reliably measure whether a decay event occurred. In a typical ion trap experiment, the efficiency of detecting a single photon is very low. Fortunately, the scheme described in Sec.~\ref{sec:example} does not rely on detecting a single photon. The key is to have $\left|\uparrow\right\rangle$ decay into $|a\rangle$ instead of $\left|\downarrow\right\rangle$. To determine whether a decay event occurred, one measures the population in $|a\rangle$ by continuously exciting it to $|P_{1/2},F=0\rangle$. If the atom is in $|a\rangle$, it will scatter many photons; otherwise, it will not scatter any photons. This way, one can efficiently read out the population in $|a\rangle$ and thereby determine whether a decay event occurred, even though the single-photon efficiency is low. In other words, instead of detecting the single photon created during the decay from $\left|\uparrow\right\rangle$, one can just measure the many photons that are scattered when the atom is in $|a\rangle$.

How high should the readout fidelity be to observe tachyonic behavior? If there is a readout error, it is similar to an optical pumping error discussed in Sec.~\ref{sec:simulation}, i.e., one measures that there was not a decay event even though there really was. According to Fig.~\ref{fig:experiment}, a readout fidelity of 99.5\% is sufficient for observing that the group velocity exceeds the effective $c$. For comparison, a readout fidelity of 99.99\% has been achieved with a trapped ion \cite{myerson08}.

\subsection{Measurements}
Spinor-motion correlation is important to tachyonic dynamics [Eq.~\eqref{eq:dxdt}], so we describe a protocol to directly measure it. It is well known that single-qubit observables $\langle\sigma_{x,y,z}\rangle$  can be measured by resonance fluorescence of a cycling transition. This procedure can also be used to measure vibrational degrees of freedom by mapping them onto spinor ones. The expectation value $\langle x \sigma_z \rangle$ is measured by observing $\langle \sigma_z \rangle$ as follows. A combination of red and blue sidebands, $H_r=\eta \Omega (a \sigma_+ e^{i \phi_r}+a^\dag \sigma_- e^{-i \phi_r})$ and $H_b=\eta \Omega (a^\dag \sigma_+ e^{i \phi_b}+a \sigma_- e^{-i \phi_b})$, implements a state-dependent displacement when the phases are properly tuned. The corresponding dynamics, $U=e^{-i k x \sigma_x}$, acts on $\sigma_z$ according to $U\sigma_z U^{\dag} = \cos(k x)\sigma_z + \sin(k x)\sigma_y\equiv O$.  For $k\langle x \rangle \ll 1$, this can be approximated by $\mathbb{1} \sigma_z + k x \sigma_y$. Therefore, the derivative of the expectation value of $O$, $\partial_k \langle O \rangle$, corresponds to $\langle x \sigma_y \rangle$, which can be transformed into $\langle x \sigma_z \rangle$ with a rotation of the spinor. 


\subsection{Scalability}

In this paper, we discussed how to simulate a single Dirac tachyon using a single trapped ion. Here, we consider the possibility of generalizing the scheme to $N$ ions. Suppose there is a one-dimensional array of $N$ trapping potentials (as in a segmented trap), each with an ion inside. Due to the normal-mode structure from the Coulomb interaction, one applies lasers to induce a mode-mode coupling. Further lasers drive blue and red sidebands, and one conditions on the absence of a decay event. In this way, one implements a non-Hermitian many-body system. There are two questions. First, how large could $N$ be in a realistic experiment? Second, how large does $N$ have to be in order to be difficult to simulate on a classical computer?

For $N$ ions, $P_\text{success}\approx e^{-2m't'N}$. Suppose we set $m'=0.25$ and $t'=1$. Then for $N=5$, $P_\text{success}=0.08$. This is still reasonably high to be experimentally realistic. 


Now, to simulate such a system on a classical computer would require representing the wavefunction on a Hilbert space. The spin-1/2 degree of freedom has dimension 2. For the vibrational degree of freedom, the Hilbert space has to be truncated at some number of phonons, but there should be enough phonons to accurately represent the wavefunction. For example, a wave packet with a moderate value of $p$ would require 30 phonons. Thus, the Hilbert space for each ion has dimension 60, and the total Hilbert space has dimension $60^N$. For $N=5$, the dimension is $8\times 10^8$, which is difficult to simulate on a classical computer. (This is equivalent to having 30 spin-1/2 particles.) Thus, it is possible to reach a regime where $N$ is large enough to be difficult to simulate classically. This is due to the large number of phonons that must be tracked.

\section{Conclusion}

We have proposed a quantum simulation of tachyons with current ion-trap technology. This ability to experimentally observe tachyonic physics opens the door to many possibilities. For example, one can study how imaginary mass affects bound states in a confining potential \cite{hiller02} or the Dirac oscillator \cite{moshinsky89}. One can also consider a tachyon interacting with another tachyon or a normal particle \cite{feinberg67}. By including more ions, one can study many-body physics \cite{lee12,lee14b,lee14d}. Finally, tachyon physics in conjunction with metamaterials may lead to new applications in manipulating the propagation of light \cite{szameit11}.

\begin{acknowledgments}
We thank Peter Zoller, Tilman Pfau, and Rajibul Islam for useful discussions. T.E.L. was supported by NSF through a grant to ITAMP. We also acknowledge support from Basque Government IT472-10 and BFI-2012-322, Spanish MINECO FIS2012-36673-C03-02, Ram\'on y Cajal Grant RYC-2012-11391, UPV/EHU Project No. EHUA14/04, UPV/EHU UFI 11/55, PROMISCE and SCALEQIT European projects.
\end{acknowledgments}

\bibliography{tachyon}

\end{document}